# Assessing Combinatorial Design for Analyzing System Performance of a Computer Network

*Bestoun S. Ahmed [1], Amin S. Mohammad [2], and Hemin T. Essa [3]
*Software Engineering Research Group, Software Engineering Department,
Engineering College, Salahaddin University - Hawler (SUH)
Kirkuk Road, 44002 Eribl, Kurdistan Region*
[1] bestoon82@gmail.com, [2] kakshar@gmail.com, [3] hemin.essa@hotmail.com

**Abstract**

Combinatorial design concerns with the arrangement of a finite set of elements into patterns (subsets, words, arrays) according to specified rules. The usefulness of this design method is that the number of input combination can be reduced dramatically, but the combinatorial set covers all of them. This paper presents the application of this design method in communication networks. Communication engineers can use this novel method to generate test cases for producing a cost-effective set of experiments to recognize the factors that have the least and most impact on the system's performance. A familiar scenario is used for the experiment, and five factors with different values are chosen to qualify their effect on the network performance. The experimental set is generated using combinatorial design method, and then it is used to analyze the impact of each factor. The experiments showed the effectiveness of the method to be used for analyzing the effect of factors on the communication network.

## I. Introduction

Network performance evaluation is an activity aims to determine the effectiveness of a network system and to accurately find the correct and fair configuration that produce better performance than other configuration [1, 2]. Performance evaluation and analyzing gains more interest recently as researchers focused more on comparing alternative system architecture solution and protocols to investigate the better performance networks. In fact, different factors could affect the system performance ranging from protocols, channel capacity, network size, to the transmission range [2].

Computer network performance can be evaluated and analyzed either by workload characterization, analytic models, or simulated models [3]. More recent evidence shows that the interaction of multiple configuration parameters may also affect the performance [4]. To this end, in addition to the methods mentioned above, Design of Experiments (DOE) has been used to aid the performance evaluation and analyzing. The DOE concept for network analysis is to identify the system components first and then generate different configuration experiments base on the chosen rule to sample the components' values statistically. In fact, this process is used for large systems and when different configurations exist because the process is limited by cost as the addition of each experiment leads to additional expenditures. In this case, the sampling process of DOE is essential.

Based on its effectiveness and usefulness in software testing field (e.g., [5-7]), combinatorial design and optimization have been used in different fields as a sampling technique (e.g. [8-10]). It is proved through different research that it could be used effectively as another alternative for the experiment. To this end, different research nowadays seeking to investigate the application of combinatorial interaction design in various fields starting from biology, chemistry, computer architecture, software testing, design and analysis, control engineering, to others non investigated application. In line with this approach, this paper presents a new application area for the combinatorial design and optimization. Hence, the work concerned with the understanding of the relationship between the factors that affect the performance when a computer system has different configuration factors. The work shows how the combinatorial design can be used to analyze the performance of a network and lead to more sensible choices over a wide range of network conditions. Network engineers can use this novel technique to produce a cost-effective set of experiments to recognize the factors that affect the system performance.

* Corresponding Author

The rest of this paper is organized as follows. Section II gives an overview of the statistical design of experiments and its background. Section III illustrates the combinatorial interaction design, its notations, and how it is constructed. Section IV describes the system model that is used as an application for the current approach. Section V presents and discusses the simulation results of the approach followed by concluding remarks in section VI.

## II. Statistical Design of Experiments (DOE)

DOE has emerged from the past decade. Different methods have emerged in this direction. Generally, in the DOE the system is represented as components which are called "factors." Here, the test cases are called "experimental runs." The experimental run comprehends the system components which are represented by its valid configuration and values [11]. The "full factorial" design notation is used when all possibilities of test cases (i.e., Exhaustive) is considered. [11].

However, in reality, the full factorial design is not possible when the number of factors is huge, and the system is large. In this case, to reduce the experimental runs, the "fractional factorial" design is used to sample a subset of the full design. While this DOE method is used with the numerical factors, it is not used with categorical factors [12]. For this reason, the D-Optimality design is emerged to be used with the specific factors. Evidence showed that this method is more useful for reducing the experimental run [13, 14]. This because the D-Optimality selection and sampling process from the full factorial is more effective and systematical than other methods which depend on random selection. Hence the selected experimental runs are closer to the full factorial design [14].

## III. Combinatorial Design

Combinatorial design technique has been used successfully for the approximation of full factorial design as a sampling technique [14]. As in case of DOE methods, combinatorial design technique models the system under test as a set of factors for each of which has different values. In contrast with the DOE techniques, combinatorial design samples the set of inputs base on specific coverage criteria. The criteria impose to the generated set to include a particular combination of the factors [10]. Hence, for the case of the pairwise combination, it is essential to cover all the combination of two input factors in the experiment. Empirical evidence showed that combinatorial design could produce better results than full factorial approximation experiments when the D-Optimality is considered [13, 15].

Combinatorial design searches for best solution from a finite set of feasible solutions. For covering all the combinations, it is essential to cover all of them at least once. Mathematically, Covering Array (CA) has been introduced to represent all those combinations. A $CA_\lambda$ (N;t,k,v) represents an N×k array with $v$ values such that every N×t sub-array contains all ordered subsets from $v$ values of size $t$ at least λ times [16] where k is the number of components (parameters). For optimal combination-set, we usually want all $t$-combinations to occur at least once. In this case, we consider the value of λ=1, and the notation becomes CA (N;t,k,v) [17]. As we are searching for the optimal set, the size of N, which is the size of the combination-set has to be as minimal as possible.

Although it has been applied successfully in few fields of science, applying combinatorial design in practice is difficult. It has been successful in hardware testing [18], gene expression regulation [19], advance material testing [20], control system engineering and PID tuning [9], and different software testing applications [6, 7, 21, 22]. However, as suggested by [10], the direction of research recently has been shifted to find the application of combinatorial design for different input domains in different fields. Hence, the goal of this research is to find a new application of it in a new field which is computer network performance evaluation and analyzing.

## IV. System Model

To test the applicability of the combinatorial design method, we choose a computer network scenario. The network of the scenario is running EIGRP [23] as a routing protocol on every interface. The topology of the network consists of seven connected routers, with one application server and a client. Figure 1 illustrates the topology of the used scenario.

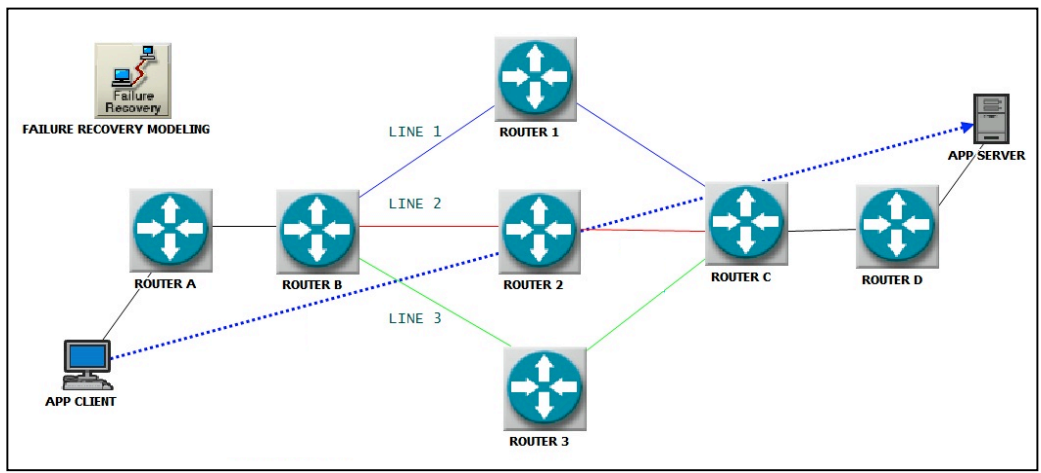

Figure 1. The topology of the used scenario

As all the routers use EIGRP protocol, this research considered five factors in this topology. These factors are, Load Balancing, TCP Parameter, Hello Interval Time, IP Forwarding Class, and Receive Buffer.

The load balancing factor attempts to distribute the traffic over various paths to optimize the network utilization [24]. The factor accepts two values which are the destination- based and packet-based. In the former one, all packets that belong to the same traffic flow are identified by the source-destination IP address par, and they are guaranteed to flow within the same route to the destination. While the packet-based value, distributes the traffic on a pre-packet basis and the successive packets are forwarded on alternative paths to the destination where the paths are selected in a round robin fashion [25].

TCP parameter can take several attributes. Here, the attributes "availability of fast recovery feature" and "receive buffer Size" are chosen. For the availability of fast recovery attribute, the predefined values by simulator are chosen (i.e., Reno, New Reno, and Disable); each of which indicating that the node is configured to support one of them. Both Reno and Default settings of TCP Parameters attributes correspond to the same TCP configuration that implements the Reno flavor of TCP. TCP New Reno value is an improvement over TCP Reno because the inefficiency related to multiple losses within a single window of data is removed [26].

Hello Interval specifies the interval between two consecutive Hello messages. The value should be the same for all interfaces connected to the same network. The default value for this attribute is 10 seconds for all interface types except for MANET interfaces, which is set to 2 seconds by default.

IP forwarding class (also called Class of Service CoS) represents a configuration in IP QoS parameters of the router's interface. The forwarding class is essential in the case when there is a relationship with other forwarding classes. The forwarding class represents a method to weigh the relative importance of one packet over another in a different forwarding class

Receive Buffer is configured to have multiple values. The attribute Receive Buffer (bytes) specifies the total amount of space available at the receiver to store arriving data before its forwarding to the upper layers [27]. For the Receive Buffer Size, three values are considered in addition to the default setting. These values are 8760, 32768, and 65535.

The factors' values are identified during the establishment of the model. Table 1 summarized the factors that have been used within our model and their values. Note that to run the full experiment of these factors, it is essential to run (2×3×4×4×4) experiment which equal to 384 experiments.

Table 1: System model factors and configurations

|  | Factors | | | | |
|---|---|---|---|---|---|
|  | Load Balancing | TCP parameter-Fast recovery | Hello Interval Time | IP forwarding Class | Receive Buffer(bytes) |
| Configurations or Values | Base on Packets | Reno | 5 | best-effort | 8760 |
|  | Base on Destination | New Reno | 10 | expedited-forwarding | 32768 |
|  |  | Disable | 15 | assured-forwarding | 65535 |
|  |  |  | 3 | network -control | default |

## V. Simulation Results and Discussion

The OPNET Simulation tool is used as a simulator to simulate the network. OPNET is a dedicated tool for network design as a finite state machine model. It can model protocols, devices, and behaviors. The OPNET IT Guru academic edition is used for this research.

After identifying the factors and their corresponding values as shown in Table 1, they are used with the combinatorial strategy to construct the combinatorial set of experiment. To construct the combinatorial set, our previous developed strategy for construction is used in [7, 21]. Table 2 shows the constructed combinatorial set of experiments. Using the strategy, 16 cases have been built systematically in which each case in the table represents a configuration for the network. Hence, the 384 possibilities of experiments are summarized in 16 experiments.

Table 2: Experiments used with the model to analyze the system

| Exper. # | Load_Balancing | TCP_parameter | Hello_Interval_Time | IP_forwarding_Class | Receive_Buffer |
|---|---|---|---|---|---|
| 1 | Based on Destination | New Reno | 5 | best-effort | 32768 |
| 2 | Based on Packets | Disable | 5 | expedited-forwarding | 65535 |
| 3 | Based on Destination | Reno | 5 | assured-forwarding | default |
| 4 | Based on Packets | New Reno | 5 | network-control | 8760 |
| 5 | Based on Packets | Reno | 10 | best-effort | 65535 |
| 6 | Based on Destination | New Reno | 10 | expedited-forwarding | default |
| 7 | Based on Destination | Disable | 10 | assured-forwarding | 8760 |
| 8 | Based on Destination | Reno | 10 | network-control | 32768 |
| 9 | Based on Packets | Disable | 15 | best-effort | default |
| 10 | Based on Destination | Reno | 15 | expedited-forwarding | 8760 |
| 11 | Based on Packets | New Reno | 15 | assured-forwarding | 32768 |
| 12 | Based on Destination | New Reno | 15 | network-control | 65535 |
| 13 | Based on Packets | Reno | 3 | best-effort | 8760 |
| 14 | Based on Destination | Disable | 3 | expedited-forwarding | 32768 |
| 15 | Based on Destination | New Reno | 3 | assured-forwarding | 65535 |
| 16 | Based on Destination | Disable | 3 | network-control | default |

As shown in Table 2 each experiment represents a configuration which is the combination of 5 values. Each of them is applied to the model's simulation to draw its performance. Packet drop is considered to evaluate the performance of each designed experiment. Figure 2 shows the aggregate number of dropped packets in intervals of duration 11h for the 16 experiments.

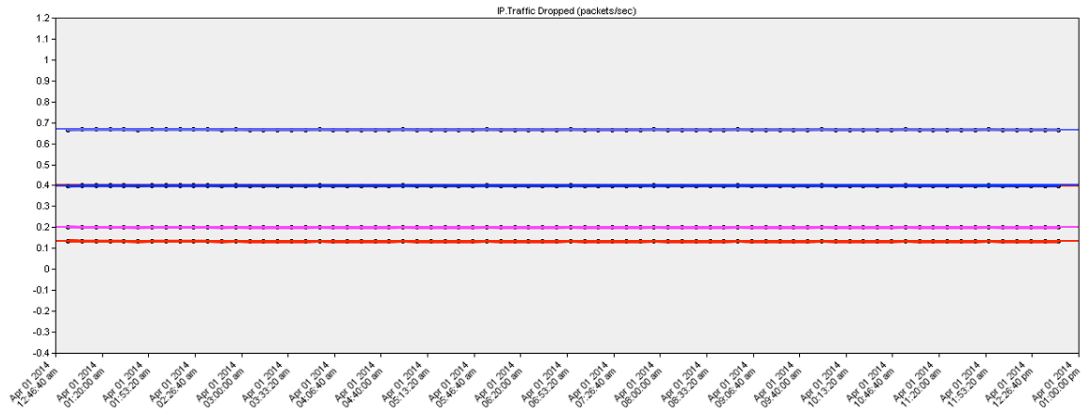

Figure 2. Aggregate number of dropped packets in intervals of duration 11$_h$ for 15 test cases

As can be seen from Figure 2, the performance of the 16 experiments is separated into four groups. The first group where the dropped packets are more than 0.65 packets/s; the second group where the dropped packets are about 0.4 packets/s; the third group where the dropped packets are about 0.2 packets/s; and fourth group where the dropped packets are more than 0.1 packets/s. By analyzing and evaluating the result, it is observed that the worst performance values were measured within the first group of results where the experiment 12, 13, 14 and 15 in Table 2 executed (see Figure 3). The experiments also indicate that the 4th group which consists of experiments 9, 10, and 11 has the best performance value. Figure 4 shows the number of dropped packets for the experiments 9, 10, and 11 in which the best performances were observed.

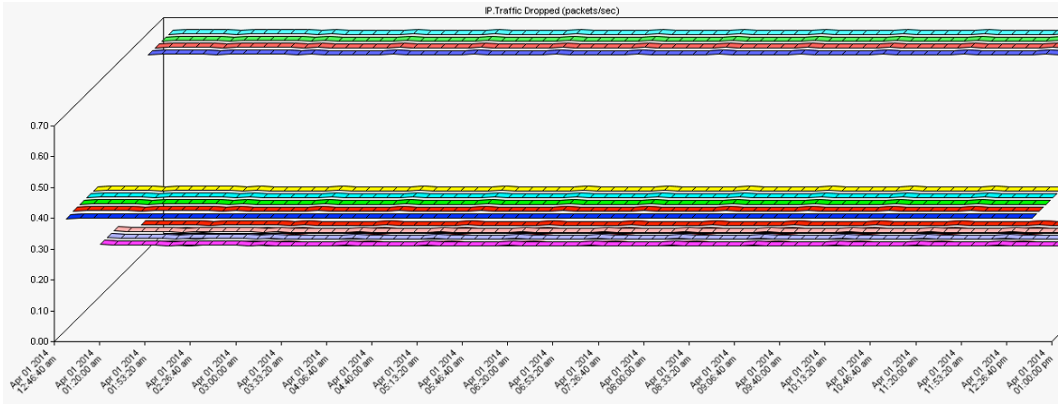

Figure 3. Number of dropped packets Vs Time with the indication of experiment cases

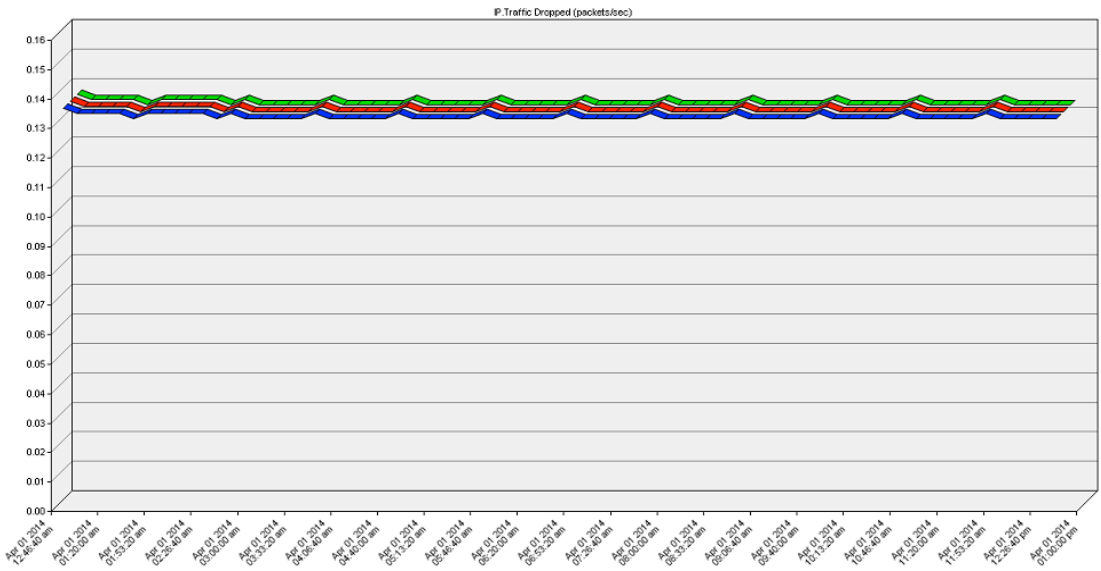

Figure 4. Packet loss distribution for experiments 9, 10 and 11

The results indicated that the best performance could be observed when the best combination is chosen. As mentioned, the experiments 9, 10, and 11 observed better performance. Each experiment of them consists of a network configuration. Using the combinatorial design it is clear that the better performance can be observed when the combination of *<Based on Packets, Disable, 15, best-effort, default>*, *<Based on Destination, Reno, 15, expedited-forwarding, 8760>* or *<Based on Packets, New Reno, 15, assured-forwarding, 32768>* are chosen. It is also clear from the experiments that the best performance can be observed within the combinations only when the *Hello_Interval_Time* is 15.

## VI. Conclusion

This paper aims to point out a novel application approach to combinatorial design in computer network system. The new approach proved its ability to detect the best configurations that get the best performance within 16 experiments (see Table 2). Within these 16 experiments, the approach indicates that three configurations could observe the better performance as compared to the other 16 configurations. However, to run the full experiment, there are 384 possible experiments for running. The new approach could be applied in many different communication systems in the same way.

## Acknowledgement

This research is partially supported by the IT center of Engineering College, Salahaddin University - Hawler (SUH).